\documentclass[aps,pre,reprint,amsmath,amssymb]{revtex4-2}

\usepackage{amsmath,amssymb,amsfonts,bm}
\usepackage{graphicx}
\usepackage{hyperref}
\usepackage{xcolor}
\usepackage{float}
\usepackage{physics}
\usepackage{overpic}
\usepackage{tikz}
\usepackage{float}

\usetikzlibrary{arrows.meta, positioning, shapes, decorations.pathmorphing}

\hypersetup{
    colorlinks=true,
    linkcolor=blue,
    citecolor=blue,
    urlcolor=blue,
    pdftitle={CMGP PRE Manuscript},
    pdfauthor={}
}

\begin{document}

\title{The Memory Engine: Self-Organized Coherence from Internal Feedback}

\author{Aranyak Sarkar\thanks{Email: \href{mailto:Aranyak.Sarkar@uni-bayreuth.de}{Aranyak.Sarkar@uni-bayreuth.de}, \href{mailto:aranyak@barc.gov.in}{aranyak@barc.gov.in}}}

\affiliation{Experimental Physics I, University of Bayreuth, Universitätsstr. 30, 95447 Bayreuth, Germany}
\affiliation{Radiation \& Photochemistry Division, Bhabha Atomic Research Centre, Mumbai 400085, India}

\begin{abstract}
We present a continuous-space realization of the Coupled Memory Graph Process (CMGP), a minimal non-Markovian framework in which coherence emerges through internal feedback. A single Brownian particle evolves on a viscoelastic substrate that records its trajectory as a scalar memory field and exerts local forces via the gradient of accumulated imprints. This autonomous, closed-loop dynamics generates structured, phase-locked motion without external forcing. The system is governed by coupled integro-differential equations: the memory field evolves as a spatiotemporal convolution of the particle’s path, while its velocity responds to the gradient of this evolving field. Simulations reveal a sharp transition from unstructured diffusion to coherent burst--trap cycles, controlled by substrate stiffness and marked by multimodal speed distributions, directional locking, and spectral entrainment. This coherence point aligns across three axes: (i) saturation of memory energy, (ii) peak transfer entropy, and (iii) a bifurcation in transverse stability. We interpret this as the emergence of a \textit{memory engine}---a self-organizing mechanism converting stored memory into predictive motion---illustrating that coherence arises not from tuning, but from coupling.

\end{abstract}

\maketitle

\section*{Introduction}

Coherent patterns often emerge in noisy systems—from intracellular transport to biological swarms and soft robotics—despite fluctuations at the microscale~\cite{Gammaitoni1998,Pikovsky2001,Lindner2004}. Classical explanations attribute such order to external forcing, global coupling, or spatial constraints. Yet, many natural and synthetic systems display self-organization without centralized control, where order arises from local interactions mediated by an evolving environment~\cite{Lauga2009,Camley2017,Rafsanjani2019}.

Recent experiments exemplify the principle of memory-mediated or environment-coupled feedback. In walking droplet systems, a bouncing particle generates a persistent wave field that guides its own trajectory, enabling quantized orbits and coherent motion~\cite{Couder2006,Fort2010, Perrard2014}. When this wave field saturates, responsiveness diminishes—leading to memory-driven amnesia~\cite{Hubert2022}.

A key mechanism in such systems is feedback between motion and history: agents modify their surroundings, which in turn influence their future dynamics. This closed-loop causality—where memory is both written and re-read—challenges traditional non-Markovian models~\cite{Cross1993,Scholl2008,Ramaswamy2010}, which typically treat memory as an externally imposed, static influence. Nonetheless, delayed and structured interactions have been shown to produce coherence without global coordination~\cite{Cross1993,Scholl2008,Ramaswamy2010}.

While memory has been incorporated through generalized Langevin equations and nonlocal kernels~\cite{Kubo1966,Ramaswamy2010}, these approaches describe memory passively—as a fixed kernel shaping dynamics unidirectionally. In contrast, many real-world systems operate in deformable media that actively evolve in response to motion and feed information back. Examples include viscoelastic matrices that retain stress~\cite{Angelini2011}, walking droplets guided by self-generated wave fields~\cite{Couder2006}, and soft robots whose shape and motion are programmed through mechanically reconfigurable metamaterials~\cite{Rafsanjani2019}. Even in simulation-based systems, the emergence of synchrony from local memory-like interactions can be predicted from initial states via machine learning~\cite{Wang2022CollectiveGNN}. In such cases, memory is not statically imposed but dynamically constructed and re-accessed, forming a reciprocal loop. Across these diverse systems, coherence does not arise from imposed control, but through reciprocal interaction with an evolving environment—mechanical~\cite{Angelini2011,Rafsanjani2019}, hydrodynamic~\cite{Lauga2009,Couder2006}, or field-based~\cite{Fort2010,Perrard2014}. The emergence of structure from self-reinforcing feedback loops, where past motion shapes future behavior, points toward a more general design principle: that of a \textit{memory machine}—a self-driven engine that stores, reprocesses, and utilizes its own history to sustain structured dynamics. 

Here, we seek to formalize this idea by developing an optimal memory engine: one that maintains a non-dissipative flow of information while operating efficiently under stochastic fluctuations. Our goal is to isolate the minimal ingredients required for such autonomous coherence, where memory is not passively imposed but dynamically constructed and functionally reused. To this end, we introduce a theoretical framework that distills these principles into a continuous, memory-active setting.

This framework extends our Coupled Memory Graph Process (CMGP) architecture~\cite{sarkar2025nonmarkovian}—previously developed for discrete, memory-coupled agents—into continuous space. By embedding memory in a dynamic field and coupling it directly to velocity, we demonstrate that feedback alone suffices to generate structured motion such as phase locking, directional coherence, and burst--trap cycles, without any external tuning.

Recent theoretical work has begun to explore how autonomous feedback can give rise to structure in active systems. A notable example is the information engine proposed by Cocconi and Chen~\cite{Cocconi2024}, where a single run-and-tumble particle interacts with an auxiliary inference variable to extract work from thermal fluctuations. While both their framework and ours emphasize internal feedback and the reuse of memory under noise, their focus lies in maximizing thermodynamic efficiency through a purposefully designed control loop. By contrast, our model operates without external controllers or measurement protocols. Here, memory is not encoded through inference, but imprinted in a deformable field shaped by the particle’s motion—turning the environment itself into a dynamic store of history. This spatially embedded memory enables the spontaneous emergence of coherence, not as an optimized outcome, but as a natural consequence of reciprocal, history-dependent coupling between motion and environment.

In this setting, coherence arises through the alignment of three dynamical landscapes:
\begin{enumerate}
    \item An \textbf{energy landscape}, where memory injection and dissipation reach a steady balance;
    \item An \textbf{entropy landscape}, where directional transfer entropy peaks, marking maximal feedback flow;
    \item A \textbf{stability landscape}, where straight trajectories destabilize and give way to phase-locked motion.
\end{enumerate}

Together, these features position CMGP as a general framework for modeling self-organization in memory-active systems. While the present work focuses on a single particle coupled to a scalar field, the formulation readily extends to multiple interacting agents, vectorial feedback, and adaptive media. By unifying feedback, memory, and stochasticity within a closed dynamical loop, the model reveals a simple organizing principle: \textit{coherence is not imposed—it emerges from coupling}.

\section*{Theoretical Framework}

\subsection{Coupled Memory Graph Process in Continuous Space}

\begin{figure}[t]
\centering
\begin{overpic}[width=0.35\textwidth]{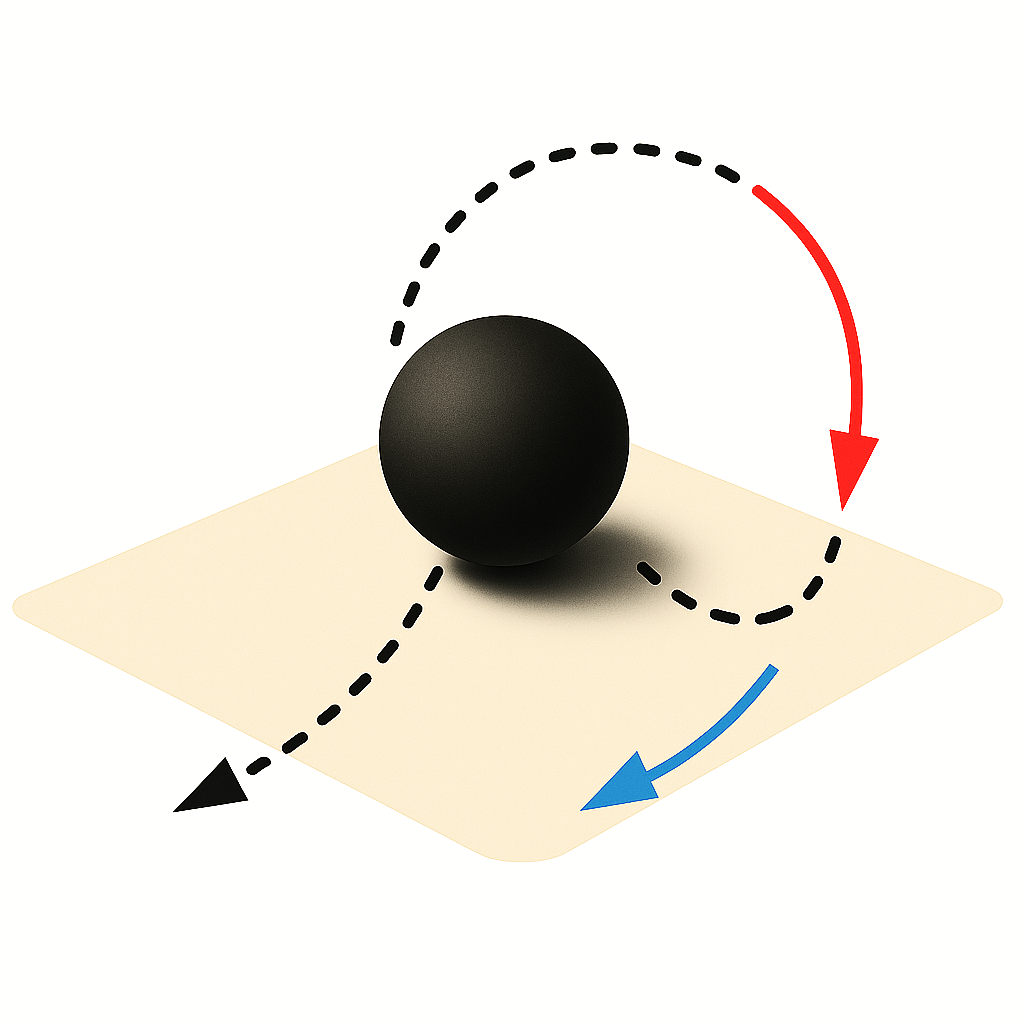}

  \put(86,73){\color{red} \scriptsize $G_{\sigma}(\mathbf{r} - \mathbf{r}(\tau))$}

  \put(35,91){\color{black} \scriptsize $\Theta(t - \tau)$}
  \put(28,58){\color{black} \scriptsize $\mathbf {r}(t)$}
  \put(68,48){\color{blue} \scriptsize $\nabla S$}

  \put(40,25){\color{black} \scriptsize $S(\mathbf{r}, t)$}

\end{overpic}
\caption{Schematic of surface-mediated memory and feedback in the CMGP model.}
\label{fig:cmgp-surface}
\end{figure}
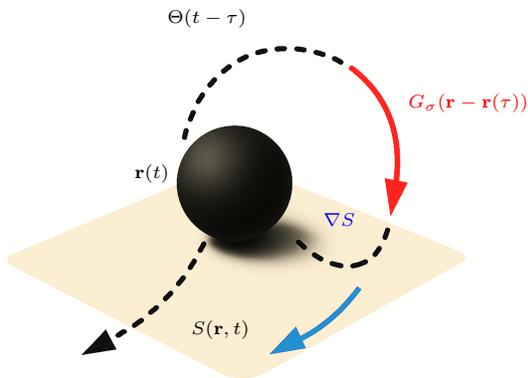


Building on our Coupled Memory Graph Process (CMGP) framework~\cite{sarkar2025nonmarkovian}---which modeled temporal coherence and non-Markovian interactions across discrete units---we now extend the paradigm to a continuous spatial setting. Here, a single memoryless Brownian particle evolves in two dimensions while interacting with a dynamically evolving scalar memory field. This field records the particle's trajectory as a decaying spatial imprint and simultaneously influences its motion through local feedback, forming a closed-loop, memory-mediated system.

Let \( \mathbf{r}(t) \in \mathbb{R}^2 \) denote the particle's position, and let \( S : \mathbb{R}^2 \times \mathbb{R}_+ \to \mathbb{R} \) represent the memory field encoding both spatial and temporal history. The particle continuously deposits localized traces into \( S \), which decay over time, shaping a history-dependent landscape. In turn, the particle's motion is guided by the gradient \( \nabla S(\mathbf{r}(t), t) \), enabling interaction with its own past as shown in Figure~\ref{fig:cmgp-surface}. This feedback loop gives rise to complex behaviors such as delayed self-attraction, phase locking, and emergent coherence.

\subsection{Memory Field Dynamics}

The surface memory scalar field evolves according to the integro-differential equation:
\begin{equation}
\partial_t S(\mathbf{r}, t) = -\alpha_s S(\mathbf{r}, t) + A \int_0^t \Theta_s(t - \tau)\, G_\sigma(\mathbf{r} - \mathbf{r}(\tau))\, d\tau,
\label{eq:memory_surface}
\end{equation}
where \( \alpha_s = 1/\tau_s \) is the memory decay rate, \( \Theta_s(t) = \alpha_s e^{-\alpha_s t} \) is a normalized exponential kernel encoding causal temporal decay, and \( G_\sigma(\mathbf{r}) \) is a spatially localized Gaussian imprint defined as
\begin{equation}
G_\sigma(\mathbf{r}) = \frac{1}{2\pi\sigma^2} \exp\left( -\frac{\|\mathbf{r}\|^2}{2\sigma^2} \right),
\label{eq:imprint}
\end{equation}
with \( \sigma \) setting the spatial resolution of memory. The parameter \( A \) has units of inverse time and controls the deposition rate.

Formally, Eq.~\eqref{eq:memory_surface} defines a spatiotemporal convolution of the particle's trajectory:
\[
S(\mathbf{r}, t) = \left[ \Theta_s *_{t} \left( G_\sigma *_{\mathbf{r}} \delta(\mathbf{r} - \mathbf{r}(\cdot)) \right) \right](\mathbf{r}, t),
\]
where \( *_{t} \) and \( *_{\mathbf{r}} \) denote convolution over time and space, respectively. This representation highlights how the field \( S(\mathbf{r}, t) \) accumulates a temporally weighted, spatially smoothed memory of the particle’s path.

The convolution operations follow the definitions:
\begin{gather*}
(\Theta_s *_{t} f)(t) = \int_0^t \Theta_s(t - \tau)\, f(\tau)\, d\tau, \\
(G_\sigma *_{\mathbf{r}} f)(\mathbf{r}) = \int_{\mathbb{R}^2} G_\sigma(\mathbf{r} - \mathbf{r}')\, f(\mathbf{r}')\, d\mathbf{r}'.
\end{gather*}

Thus, the substrate is not passive—it is continuously written to and read from by the particle, forming a \emph{self-induced memory landscape}. This dynamic field acts as an internal state that encodes trajectory history, breaks detailed balance, and enables structured behavior through delayed self-interaction.

Unlike our earlier CMGP formulation~\cite{sarkar2025nonmarkovian}, where memory modulated the stochastic forcing term, we now embed memory directly into the deterministic dynamics by coupling the velocity to the particle’s own past. This transition from noise-weighted memory to history-dependent velocity convolution enables adaptation in continuous space and supports field-mediated feedback as a driver of coherent behavior. The memory field remains central, now functioning not just as a passive recorder but as an active mediator of self-interaction.

\subsection{Particle Dynamics and Feedback Loop}

The particle’s motion is governed by a nonlinear, stochastic Volterra equation of the second kind~\cite{Gripenberg1990,Pruess1993}:
\begin{equation}
\dot{\mathbf{r}}(t) = \int_0^t \Theta_m(t - \tau)\, \dot{\mathbf{r}}(\tau)\, d\tau - \kappa \nabla S(\mathbf{r}(t), t) + \boldsymbol{\xi}(t),
\label{eq:r_operator}
\end{equation}
where \( \Theta_m(t) = \alpha_m e^{-\alpha_m t} \) is the intrinsic velocity memory kernel, and \( \boldsymbol{\xi}(t) \) represents Gaussian noise. For viscoelastic or fractional systems, \( \boldsymbol{\xi}(t) \) may be generalized to colored noise~\cite{Kubo1966,Zwanzig2001}. In contrast to classical generalized Langevin equations~\cite{Kupferman2004}, our formulation places memory entirely in the deterministic part, isolating its role in structuring motion rather than modulating fluctuations.

Together, equations~\eqref{eq:memory_surface} and~\eqref{eq:r_operator} define a closed, bidirectionally coupled non-Markovian system: the particle imprints its trajectory into a deformable memory field and is steered by the resulting gradient. This autonomous feedback loop gives rise to complex dynamics such as delayed self-attraction, orbital confinement, and long-lived coherence.

In viscoelastic media, the model parameters acquire clear physical interpretations: the feedback strength \( \kappa \) scales with substrate stiffness \( E \), while the memory decay rate \( \alpha_s = 1/\tau_s \sim E/\eta \) follows from Maxwell-type relaxation with \( \tau_s \sim \eta/E \). These relations are consistent with microrheological characterizations~\cite{Ferry1980,Mason1995,Waigh2005}, where particle fluctuations reveal underlying mechanical response.
\subsection{Energy-Based Description: Memory Saturation}
To quantify the buildup of structure in the memory field, we define its instantaneous energy as the squared \( L^2 \)-norm:
\begin{equation}
\varepsilon_s(t) = \frac{1}{2} \int_{\mathbb{R}^2} S(\mathbf{r}, t)^2 \, d\mathbf{r}.
\label{eq:field_energy}
\end{equation}
Differentiating with respect to time and substituting Eq.~\eqref{eq:memory_surface}, we obtain the energy evolution:
\begin{align}
\frac{d}{dt} \varepsilon_s(t) &= -\alpha_s \int_{\mathbb{R}^2} S^2(\mathbf{r}, t)\, d\mathbf{r} \nonumber \\
&\quad + A \int_{\mathbb{R}^2} S(\mathbf{r}, t) \left[ \int_0^t \Theta_s(t - \tau)\, G_\sigma(\mathbf{r} - \mathbf{r}(\tau))\, d\tau \right] d\mathbf{r}.
\label{eq:energy_dynamics}
\end{align}
We define the dissipation and injection rates as
\begin{align}
\mathcal{D}(t) &= \alpha_s \int_{\mathbb{R}^2} S^2(\mathbf{r}, t)\, d\mathbf{r}, \label{eq:dissipation} \\
\mathcal{I}(t) &= A \int_{\mathbb{R}^2} S(\mathbf{r}, t) \left[ \int_0^t \Theta_s(t - \tau)\, G_\sigma(\mathbf{r} - \mathbf{r}(\tau))\, d\tau \right] d\mathbf{r}, \label{eq:injection}
\end{align}
so that the net energy balance reads:
\begin{equation}
\frac{d}{dt} \varepsilon_s(t) = \mathcal{I}(t) - \mathcal{D}(t).
\label{eq:net_energy_balance}
\end{equation}
Here, \( \mathcal{D}(t) \) quantifies loss due to field relaxation, while \( \mathcal{I}(t) \) measures the injection of structure from the particle’s past activity. This balance encodes the competition between memory dissipation and reinforcement, governing transitions between incoherent and self-organized regimes.

A useful analogy for this feedback mechanism is provided by walking droplet experiments, where a millimetric oil droplet bounces on a vertically oscillated fluid bath and generates waves that persist and steer its future motion~\cite{Couder2006,Bush2015}. The droplet co-evolves with its own wave field, creating a self-generated memory landscape that guides its trajectory—a classical realization of pilot-wave behavior. Our model extends this idea to a stochastic regime: instead of physical waves, the particle interacts with a deformable memory field \( S(\mathbf{r}, t) \) that encodes a smoothed, decaying trace of its past. This field deforms in response to motion and feeds back via gradient forces, forming a closed loop of structured self-interaction. Unlike a thermal bath that injects unstructured noise, the memory field regulates motion through history-dependent feedback, producing non-Markovian, adaptive dynamics that interpolate between diffusion and coherent transport.

In the steady state, \( \frac{d}{dt} \varepsilon_s(t) \to 0 \), indicating saturation—neither absorbing nor dissipating significant energy. This plateau suggests not just energetic balance, but a functional shift: from an accumulating medium to a coherent, predictive substrate that actively shapes the particle’s dynamics.

\subsection{Entropy-Based Description: Transfer Entropy}

To characterize the shift from passive storage to active influence, we transition from an energy-based to an entropy-based description. Specifically, we use \emph{transfer entropy} (TE)~\cite{schreiber2000measuring}, which quantifies the directional predictability between two stochastic processes. For time series \( X(t) \) and \( Y(t) \), the TE from \( X \) to \( Y \) is defined as
\begin{equation}
\mathrm{TE}_{X \to Y} = \sum p(y_{t+1}, y_t^{(k)}, x_t^{(l)}) \log \frac{p(y_{t+1} \mid y_t^{(k)}, x_t^{(l)})}{p(y_{t+1} \mid y_t^{(k)})},
\label{eq:te_def}
\end{equation}
where \( x_t^{(l)} \) and \( y_t^{(k)} \) denote past histories of \( X \) and \( Y \), respectively. In our system, we evaluate two directions:
\begin{itemize}
    \item \( \mathrm{TE}_{s \to p} \): how well the field gradient \( \nabla S(\mathbf{r}(t), t) \) predicts the particle’s velocity,
    \item \( \mathrm{TE}_{p \to s} \): how well the particle’s trajectory predicts the evolving field at its position.
\end{itemize}

As the substrate stiffness \( E \) or feedback strength \( \kappa \) increases, the asymmetry
\[
\Delta \mathrm{TE} = \mathrm{TE}_{s \to p} - \mathrm{TE}_{p \to s}
\]
peaks, indicating a regime where the field becomes the dominant causal agent. In this limit, the particle’s role in generating new memory diminishes, while the field gains predictive control. The alignment of energy saturation with maximal directional information flow signals an optimized feedback regime—where coherence is encoded not only energetically, but also causally.

This thermodynamic and information-theoretic framework reveals how memory saturation and causal asymmetry correlate with the onset of coherence. Yet it does not fully explain the underlying mechanism: how does a freely diffusing particle, initially governed by noise, spontaneously break symmetry and transition to structured motion—such as looping, confinement, or locking—without external forcing? To answer this, we must go beyond ensemble statistics and examine how the feedback loop dynamically reshapes randomness into coherence at the microscopic level.
\subsection{Linear Stability Analysis and Feedback Bifurcation}

To examine how coherence emerges dynamically from random initial conditions, we analyze the stability of straight-line trajectories under small transverse perturbations. This reveals whether memory-mediated feedback suppresses deviations—preserving diffusion—or amplifies them, leading to curvature, looping, or directional locking. Such analysis identifies thresholds where uniform motion becomes unstable, marking a dynamical bifurcation~\cite{strogatz2018nonlinear} in phase space, where symmetry breaks and stochastic motion becomes guided by memory.

We consider a particle moving with constant velocity \( \mathbf{v}_0 \), and perturb its trajectory orthogonally:
\begin{equation}
\mathbf{r}(t) = \mathbf{v}_0 t + \delta y(t)\, \hat{\mathbf{n}}, \qquad \|\delta y(t)\| \ll 1,
\end{equation}
where \( \hat{\mathbf{n}} \perp \mathbf{v}_0 \) is a unit transverse vector. Substituting into the equation of motion Eq.~\eqref{eq:r_operator} and linearizing in \( \delta y(t) \) gives
\begin{equation}
\delta \dot{y}(t) = \int_0^t \Theta_m(t - \tau)\, \delta \dot{y}(\tau) \, d\tau - \kappa\, \hat{\mathbf{n}} \cdot \nabla S(\mathbf{r}(t), t).
\label{eq:delta_y_dynamics}
\end{equation}
Note that Eq.~\eqref{eq:delta_y_dynamics} omits the stochastic forcing term present in Eq.~\eqref{eq:r_operator}, as the bifurcation structure emerges from deterministic feedback alone. Stochastic effects are therefore excluded without affecting the predicted instability threshold.

Now expanding \( \nabla S \) along the unperturbed path:
\begin{align}
\nabla S(\mathbf{r}(t), t) &\approx \int_0^t \Theta_s(t - \tau) \left[ \nabla G_\sigma(\mathbf{v}_0 (t - \tau)) \right. \nonumber \\
&\left. \quad + \frac{\partial}{\partial \hat{n}} \nabla G_\sigma(\mathbf{v}_0 (t - \tau)) \cdot (\delta y(t) - \delta y(\tau)) \right] d\tau.
\end{align}
Projecting along \( \hat{\mathbf{n}} \) cancels the first term (by symmetry), yielding:
\begin{equation}
\hat{\mathbf{n}} \cdot \nabla S(\mathbf{r}(t), t) \approx \int_0^t \Lambda(t - \tau)\, (\delta y(t) - \delta y(\tau))\, d\tau,
\label{eq:projected_gradient}
\end{equation}
with the effective feedback kernel defined by
\begin{equation}
\Lambda(s) := \Theta_s(s)\, \hat{\mathbf{n}}^\top \cdot \frac{\partial}{\partial \hat{n}} \nabla G_\sigma(\mathbf{v}_0 s) \cdot \hat{\mathbf{n}}.
\label{eq:Lambda_kernel}
\end{equation}

Substituting Eq.~\eqref{eq:projected_gradient} into Eq.~\eqref{eq:delta_y_dynamics} gives the closed perturbation dynamics:
\begin{align}
\delta \dot{y}(t) &= \int_0^t \Theta_m(t - \tau)\, \delta \dot{y}(\tau)\, d\tau \nonumber \\
&\quad - \kappa \int_0^t \Lambda(s)\, (\delta y(t) - \delta y(t - s))\, ds.
\label{eq:perturbation_equation}
\end{align}

This integro-differential equation captures the competition between intrinsic velocity memory (first term) and surface-induced feedback (second term). Stability requires that perturbations decay, i.e., memory damping must outpace feedback-induced amplification.

To assess this, we perform a Laplace transform \( \tilde{y}(s) = \mathcal{L}\{\delta y(t)\} \), yielding:
\begin{align}
\mathcal{L} \left\{ \int_0^t \Theta_m(t - \tau)\, \delta \dot{y}(\tau) \right\} &= \tilde{\Theta}_m(s)\, s\, \tilde{y}(s), \\
\mathcal{L} \left\{ \int_0^t \Lambda(s)\, (\delta y(t) - \delta y(t - s))\, ds \right\} &= [1 - \tilde{\Lambda}(s)]\, \tilde{y}(s).
\end{align}
Substituting into Eq.~\eqref{eq:perturbation_equation} gives the dispersion relation:
\begin{equation}
s \left[ 1 - \tilde{\Theta}_m(s) \right] + \kappa \left[ 1 - \tilde{\Lambda}(s) \right] = 0.
\label{eq:dispersion}
\end{equation}

Assuming exponential kernels:
\[
\Theta_m(t) = \alpha_m e^{-\alpha_m t}, \quad \Theta_s(t) = \alpha_s e^{-\alpha_s t},
\]
with \( \alpha_m = 1/\tau_m \), \( \alpha_s = 1/\tau_s \), and defining
\[
\Lambda(t) = \alpha_s \mathcal{K} e^{-\alpha_s t}, \quad \text{where} \quad \mathcal{K} = \hat{\mathbf{n}}^\top \cdot \partial_{\hat{n}} \nabla G_\sigma(\mathbf{0}) \cdot \hat{\mathbf{n}},
\]
we obtain Laplace transforms:
\[
\tilde{\Theta}_m(s) = \frac{1}{1 + s \tau_m}, \quad \tilde{\Lambda}(s) = \frac{\alpha_s \mathcal{K}}{1 + s \tau_s}.
\]

Substituting into Eq.~\eqref{eq:dispersion} gives:
\[
s \cdot \left( \frac{s \tau_m}{1 + s \tau_m} \right) + \kappa \cdot \left( \frac{1 + s \tau_s - \alpha_s \mathcal{K}}{1 + s \tau_s} \right) = 0.
\]

To determine the critical threshold, evaluate at \( s = 0 \):
\[
\lim_{s \to 0} \left[ s \cdot \left( \frac{s \tau_m}{1 + s \tau_m} \right) + \kappa \cdot \left( \frac{1 + s \tau_s - \alpha_s \mathcal{K}}{1 + s \tau_s} \right) \right] = \kappa (1 - \alpha_s \mathcal{K}).
\]

Instability occurs when this expression crosses zero, yielding the critical condition:
\[
\alpha_s > \alpha_c := \frac{1}{\mathcal{K}}.
\]

This defines the onset of feedback-driven instability: straight-line motion becomes unstable when the transverse curvature of the memory field outweighs its temporal decay. In this regime, the feedback loop amplifies deviations, destabilizing uniform motion and favoring curved, locked, or phase-coherent trajectories.

Thus coherence in this system arises not through external tuning but via a spontaneous bifurcation rooted in the structure of the memory field. When the curvature of the self-induced substrate outpaces its temporal decay, transverse fluctuations are amplified rather than suppressed—destabilizing straight-line motion and promoting curved or locked trajectories. This marks a transition from randomness to self-guided dynamics, driven entirely by internal feedback. While global metrics such as energy saturation and transfer entropy signal the onset of coherence, local stability analysis reveals its dynamical origin in the breakdown of uniform motion. Together, these energetic, informational, and dynamical signatures define a unified mechanism for memory-driven self-organization, where structure emerges from delayed self-interaction encoded in a deformable field. In the following section, we numerically test these predictions, quantify critical instability thresholds, and examine how substrate stiffness, feedback strength, and memory decay govern the emergence of coherent trajectories.

\section*{Numerical Simulations}
\setcounter{subsection}{0}
\begin{figure*}[t]
    \centering
    \includegraphics[width=\textwidth]{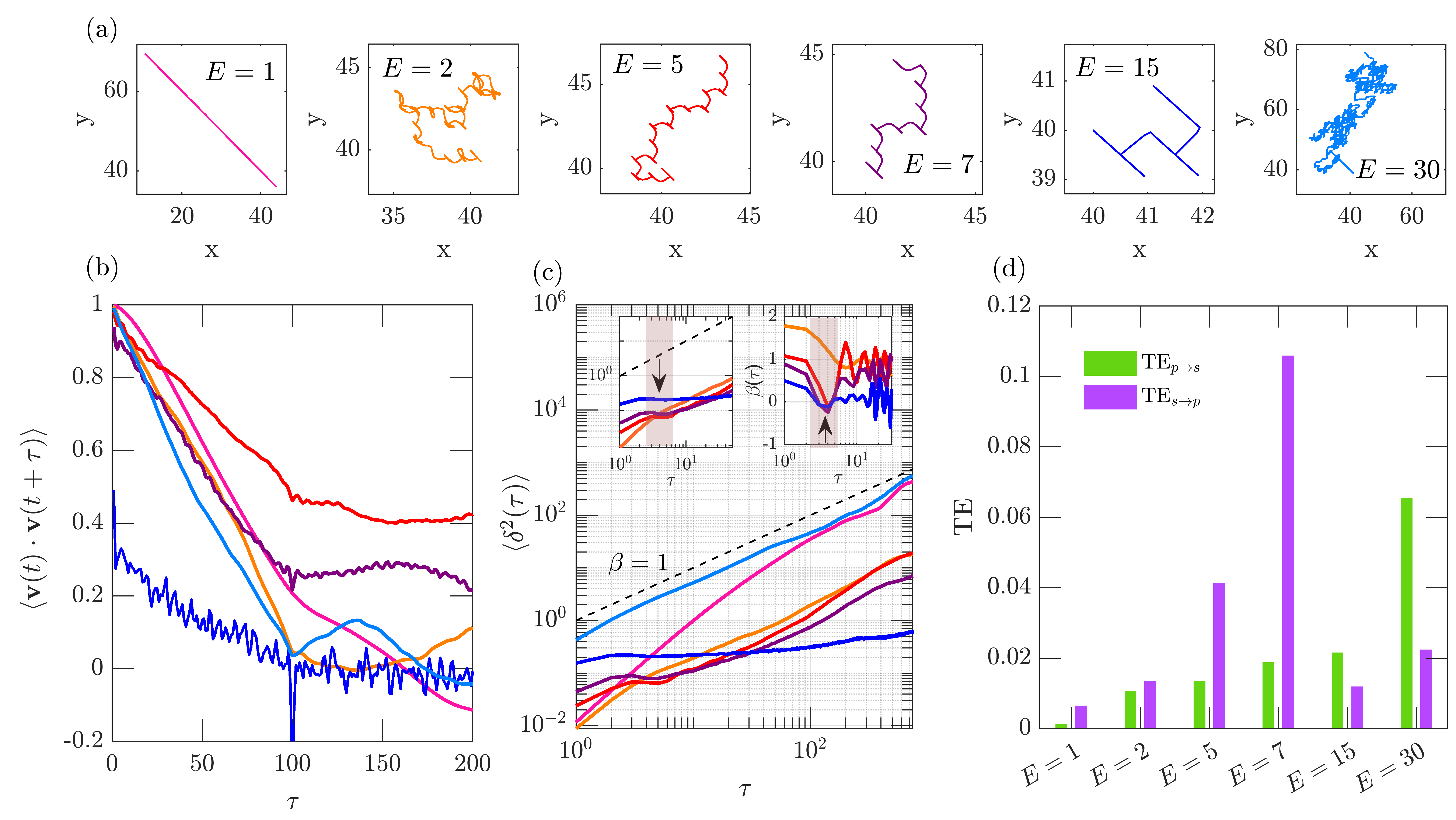}
    \caption{
\textbf{Emergence and loss of coherence with increasing substrate stiffness \( E \) at fixed viscosity \( \eta = 5 \).} 
(a) Particle trajectories illustrate a transition from smooth, viscous-dominated motion (\( E = 1 \)) to structured, phase-locked paths at intermediate stiffness (\( E = 5{-}15 \)), followed by a return to irregular motion at high stiffness (\( E = 30 \)) due to memory saturation. 
(b) Velocity autocorrelation functions (VACF) reveal enhanced temporal correlations and oscillatory behavior near the coherence regime (\( E = 5{-}7 \)), which are absent at both low and high stiffness. 
(c) Time-averaged mean squared displacement (MSD) and the corresponding dynamic exponent \( \beta(\tau) \) show a sub-to-superdiffusive crossover in the intermediate stiffness regime. The shaded region marks the feedback-induced transition window, with characteristic timescale \( \tau_c \sim \eta/E \). Oscillations in \( \beta(\tau) \) reflect confinement and release driven by feedback. 
(d) Transfer entropy analysis reveals maximal causal asymmetry at \( E = 7 \), where \( \mathrm{TE}_{s \to p} \) peaks while \( \mathrm{TE}_{p \to s} \) remains low, indicating that the surface exerts dominant predictive influence over the particle during coherent motion.
}

    \label{fig:trajectory_dynamics}
\end{figure*}
\begin{figure}[!htbp]
    \centering
    \includegraphics[width=0.9\columnwidth]{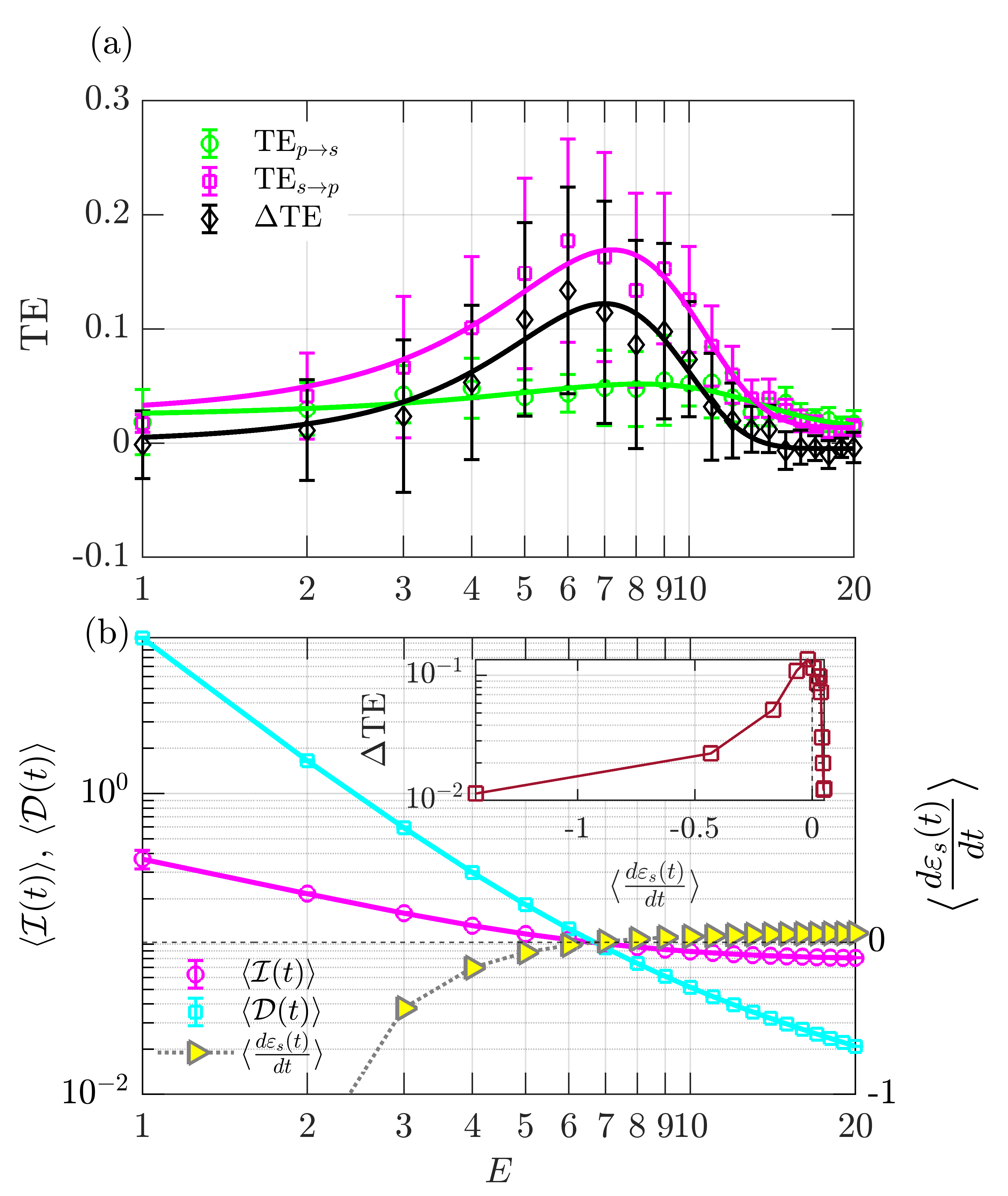}
    \caption{
\textbf{Causal and energetic signatures of coherence as a function of substrate stiffness.}
(a) Time-averaged directional transfer entropies \( \mathrm{TE}_{p \to s} \), \( \mathrm{TE}_{s \to p} \), and their asymmetry \( \Delta \mathrm{TE} \), plotted against substrate stiffness \( E \). A sharp peak in \( \Delta \mathrm{TE} \) near \( E \sim 7 \) marks the coherence window, where the memory field exerts maximal predictive influence on the particle.
(b) Mean rates of energy injection \( \langle \mathcal{I}(t) \rangle \) and dissipation \( \langle \mathcal{D}(t) \rangle \) in the memory field. The intersection of these curves aligns with the entropy peak in panel (a), identifying a self-organized balance point where memory feedback becomes maximally effective.
}

    \label{fig:energy_TE_dynamics}
\end{figure}

\begin{figure}[!htbp]
    \centering
    \includegraphics[width=\columnwidth]{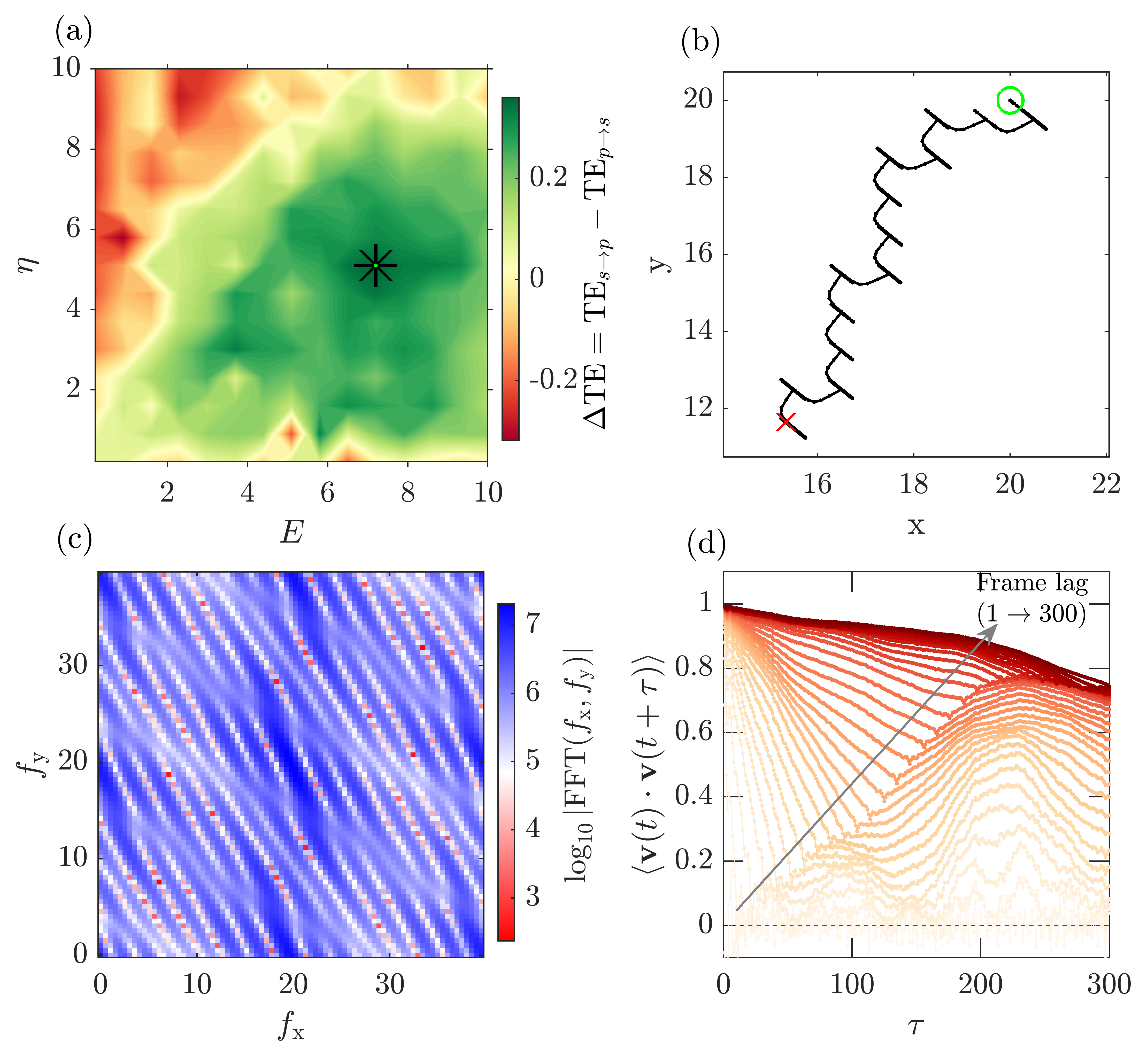}
    \caption{
\textbf{Coherence phase mapping in stiffness--viscosity space.}
(a) Heatmap of directional transfer entropy asymmetry \( \Delta \mathrm{TE} = \mathrm{TE}_{s \to p} - \mathrm{TE}_{p \to s} \) across substrate stiffness \( E \) and viscosity \( \eta \), revealing a localized hotspot near \( (7.2, 5.1) \) where feedback-driven coherence peaks.
(b) A representative trajectory from this coherence point shows staircase-like stepping and persistent directionality—signatures of structured feedback-guided dynamics.
(c) 2D Fourier transform of trajectory density reveals discrete spectral peaks, confirming spatial periodicity absent outside the coherence zone.
(d) Velocity autocorrelation function (VACF) at the coherence point exhibits long-lived oscillations, revealing rhythmic memory engagement. Together, these panels delineate a robust, memory-organized coherence regime.
}

    \label{fig:phase_map}
\end{figure}

To test our theoretical predictions, we numerically simulate the motion of a Brownian particle evolving on a viscoelastic memory substrate that records its past trajectory as decaying Gaussian-shaped imprints and exerts delayed feedback via local gradient forces as described in theory section. The particle dynamics are governed by Eq.~\eqref{eq:r_operator}, while the memory field evolves according to Eq.~\eqref{eq:memory_surface}. The imprint width is set by the particle radius, \( \sigma = R \), and the memory decay times for both field and velocity are matched as \( \tau_s = \tau_m \sim \eta / E \), yielding a unified decay rate \( \alpha = 1/\tau_s \) and feedback strength \( \kappa \sim E \sigma^2 \). All simulations are performed using non-dimensionalized units, chosen to isolate the system’s intrinsic dynamics and facilitate generality across physical realizations.

\subsection{Simulation Setup and Parameter Mapping}

The memory field is discretized on a square spatial grid and updated through exponentially weighted accumulation of Gaussian imprints. At each time step, the surface gradient is interpolated at the particle’s position and used to compute the feedback force in Eq.~\eqref{eq:r_operator}, scaled by \( \kappa \). This coupling between additive velocity memory and deformable substrate feedback enables the particle to respond to its own history in a closed-loop manner.

At the beginning, to isolate the effect of stiffness on the emergence of coherence, we fix the effective viscosity at \( \eta = 5 \), balancing memory persistence against overdamping. Simulations are conducted in a periodic domain with time step \( \Delta t = 1\), and ensemble averaging is performed over multiple independent realizations. 

To uncover how memory-driven feedback modulates motion, we systematically vary substrate stiffness \( E \) and examine how coherence emerges, stabilizes, and eventually collapses. The goal is not only to detect the presence of structure, but to trace its dynamical signatures across physical and informational observables.

\subsection{Coherence Emergence: Morphology, Dynamics and Information}

We first examine how trajectories change across increasing stiffness \( E \). Figure~\ref{fig:trajectory_dynamics}(a) shows representative trajectories as \( E \) increases. At low stiffness (\( E = 1 \)), motion is nearly ballistic, governed by velocity memory alone with negligible surface influence. As stiffness increases (\( E = 2{-}15 \)), the memory field begins to steer motion, producing curvature and closed loops. Peak coherence is observed at \( E = 7{-}15 \), where trajectories become phase-locked and recurrent. However, at higher stiffness (\( E = 30 \)), the memory field saturates rapidly, losing adaptability, and the particle reverts to disordered diffusion.

To evaluate temporal organization, we compute the velocity autocorrelation function (VACF), \( \langle \mathbf{v}(t) \cdot \mathbf{v}(t+\tau) \rangle \), which reflects the persistence of direction and speed across lag times \( \tau \) (Fig.~\ref{fig:trajectory_dynamics}(b)). In the coherence regime, the VACF decays slowly and exhibits mild oscillations, signaling long-range temporal memory. In contrast, for low and high \( E \), correlations decay quickly—either due to weak feedback or excessive memory saturation—indicating that sustained organization only exists within an intermediate mechanical window.

Memory also modifies transport behavior. In Figure~\ref{fig:trajectory_dynamics}(c), we analyze the time-averaged mean square displacement (MSD) and its dynamic exponent \( \beta(\tau) = d\log \langle \delta^2(\tau) \rangle / d\log \tau \). For intermediate stiffness, we observe a two-phase response: an initial subdiffusive regime transitions to superdiffusive growth beyond a characteristic lag time \( \tau_c \sim \eta/E \). This timescale marks the activation of memory feedback. For \( E = 5{-}7 \), oscillations in \( \beta(\tau) \) reveal alternating episodes of propulsion and slowdown—reflecting intermittent engagement with the memory field. These modulations vanish at high stiffness, where rigidity suppresses dynamical flexibility.

Finally, Figure~\ref{fig:trajectory_dynamics}(d) probes directional influence via transfer entropy. The surface-to-particle entropy flow, \( \mathrm{TE}_{s \to p} \), peaks sharply in the coherent regime, indicating that the field actively informs and constrains motion. In contrast, \( \mathrm{TE}_{p \to s} \) remains relatively flat until the last point, highlighting a strong asymmetry in the feedback loop: the field controls the particle, but not vice versa.

These observables—trajectory shape, temporal correlations, transport exponents, and directional information flow—all align within a narrow mechanical window of coherence. In this regime, memory is neither too short to inform the future nor too rigid to adapt. Instead, the system strikes a balance: memory persists just long enough to guide motion, without overwhelming it. Outside this regime, either diffusion or saturation dominates, and coherence dissolves into noise.

\subsection{Energy and Entropy Landscapes}

While Figure~\ref{fig:trajectory_dynamics} established the phenomenology of coherent motion---via trajectory morphology, temporal correlations, and transport scaling---Figure~\ref{fig:energy_TE_dynamics} shifts focus to the mechanism underpinning this coherence by analyzing two complementary landscapes: the directionality of information flow and the thermodynamic exchange of energy.

Panel~(a) quantifies the causal structure of the particle--field interaction using time-averaged transfer entropy (TE) in both directions. The forward TE, \( \mathrm{TE}_{p \to s} \), measures how much the particle’s motion predicts changes in the memory field and remains low across all stiffness values, indicating that the particle plays a passive role in shaping the substrate. In contrast, the reverse TE, \( \mathrm{TE}_{s \to p} \), quantifies how predictive the field is of the particle’s behavior. It exhibits a pronounced peak near \( E \sim 7 \), highlighting a regime where the surface memory becomes maximally informative. The resulting asymmetry \( \Delta \mathrm{TE} = \mathrm{TE}_{s \to p} - \mathrm{TE}_{p \to s} \) sharply identifies a feedback-dominated window in which the substrate takes control. This peak aligns with the phase-locked trajectories and long-lived VACF oscillations observed earlier, linking macroscopic coherence to microscopic directional predictability.

Panel~(b) examines the thermodynamic basis of this feedback loop by plotting the time-averaged rates of energy injection \( \langle \mathcal{I}(t) \rangle \) and dissipation \( \langle \mathcal{D}(t) \rangle \) within the memory field. At low stiffness, dissipation dominates injection, indicating a passive field that rapidly forgets past imprints. At intermediate stiffness, the two rates intersect, yielding a vanishing net energy flow \( \langle d\varepsilon_s/dt \rangle \approx 0 \)---a dynamically saturated regime where the field remains energetically balanced yet functionally active. Beyond this point, injection drops while dissipation continues, signaling memory saturation and inertial loss of responsiveness.

Strikingly, the inset reveals that the peak in causal asymmetry \( \Delta \mathrm{TE} \) coincides precisely with this energetic crossover. Coherence, therefore, arises not from maximal activity or dissipation but from a critical balance—where information transfer is high and energy flow is neutral. In this regime, the substrate behaves as a tuned memory reservoir: dynamic enough to steer motion, yet stable enough to avoid energetic overreach.

We interpret this balance point as a non-equilibrium fixed point—a steady state in which the particle’s motion is sustained not by continuous energy injection, but by maximal directional guidance from its own history. Coherence is thus maintained through internally regulated feedback, where motion emerges from information flow rather than force. In this regime, the system functions as an effective \textit{memory engine}: a closed-loop architecture in which stored trajectories steer future dynamics. When operating near this critical balance, the system enters a lossless, information-guided regime that we term a \textit{superinformal state}, characterized by persistent spatial and temporal coherence sustained by memory alone. The particle navigates a landscape of its own making—guided by memory, not driven by external input. All observables in Figure~\ref{fig:energy_TE_dynamics} are averaged over 3000 time steps per trajectory, confirming that the observed coherence is a robust, long-term outcome of self-organized dynamics.

\subsection{Coherence Phase Map}

To test whether coherence is a transient anomaly or a robust phase, we extend our analysis over the full \( (E, \eta) \) parameter space. Figure~\ref{fig:phase_map} maps how structural order, temporal persistence, and directional information flow co-emerge within a sharply localized region—revealing a coherence island.

Panel~(a) shows a heatmap of directional transfer entropy asymmetry \( \Delta \mathrm{TE} = \mathrm{TE}_{s \to p} - \mathrm{TE}_{p \to s} \), which quantifies the net information flow from memory surface to particle. A pronounced hotspot appears near \( (E, \eta) = (7.2, 5.1) \), indicating maximal memory-driven guidance. Beyond this point, coherence dissolves due to either weak imprinting at low \( E \), oversaturation at high \( E \), or excessive damping at high \( \eta \). The coherence zone thus reflects a narrow mechanical balance where the substrate responds and persists at just the right timescale.

Panel~(b) illustrates a representative trajectory from this coherence point. The motion exhibits staircase-like stepping and directional persistence—neither random nor ballistic. Each segment reflects a burst–trap cycle as the particle navigates its own imprint field, echoing the phase-locked patterns discussed earlier.

To quantify spatial regularity, panel~(c) presents the 2D Fourier transform of the trajectory density. Clear discrete peaks emerge, confirming underlying spatial periodicity—evidence of global ordering induced by memory-mediated feedback. Outside the coherence regime, this spectral signature collapses into isotropic noise.

Panel~(d) probes temporal order through the velocity autocorrelation function (VACF). At the coherence point, VACF exhibits long-lived oscillations, in contrast to rapid decay seen elsewhere in parameter space. These oscillations arise from recurrent feedback delays: memory is not just retained but rhythmically re-encountered.

Taken together, the four panels delineate a \textit{self-organized feedback regime} marked by spatial periodicity, temporal coherence, and maximal surface-to-particle information flow.

\begin{figure*}[!htbp]
    \centering
    \includegraphics[width=\textwidth]{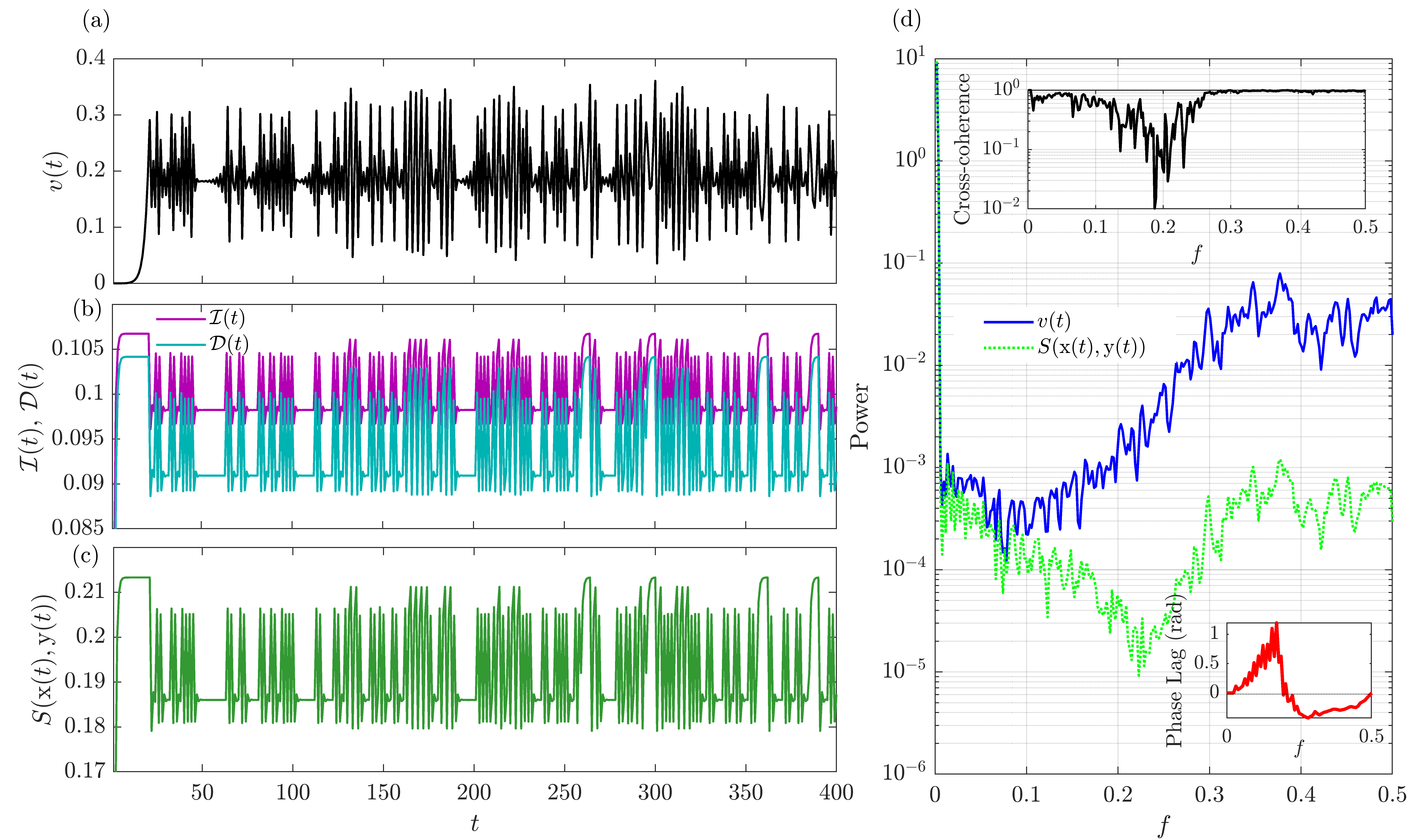}
    \caption{
\textbf{Temporal burst dynamics and spectral signatures of memory-driven coherence.}
(a) Speed trace \( v(t) \) of a representative trajectory at \( E = 7.2, \eta = 5.1 \), showing intermittent bursts of motion. 
(b) Energy injection \( \mathcal{I}(t) \) (dark purple) and dissipation \( \mathcal{D}(t) \) (green) exhibit synchronized spikes aligned with velocity bursts. 
(c) Local surface signal \( S(x(t), y(t)) \)—the memory field sampled at the particle's position—also peaks during bursts, indicating re-entry into imprinted regions. 
(d) Power spectral densities (PSDs) of particle speed (blue) and surface signal (green) reveal enhanced low-frequency power, centered around \( f \sim 0.2\). 
Top inset: cross-spectral coherence between surface and speed magnitude reveals partial entrainment, with a notable dip near \( f \sim 0.2\), indicating a loss of synchronization at this frequency. 
Bottom inset: phase lag between surface signal and speed magnitude remains relatively flat over the coherent band, suggesting consistent timing across bursts.
Together, these panels demonstrate that velocity bursts are modulated by memory-mediated feedback, generating structured, self-timed motion through a spectrally organized feedback cycle.
}

    \label{fig:burst_psd}
\end{figure*}

\begin{figure*}[t]
    \centering
    \includegraphics[width=\textwidth]{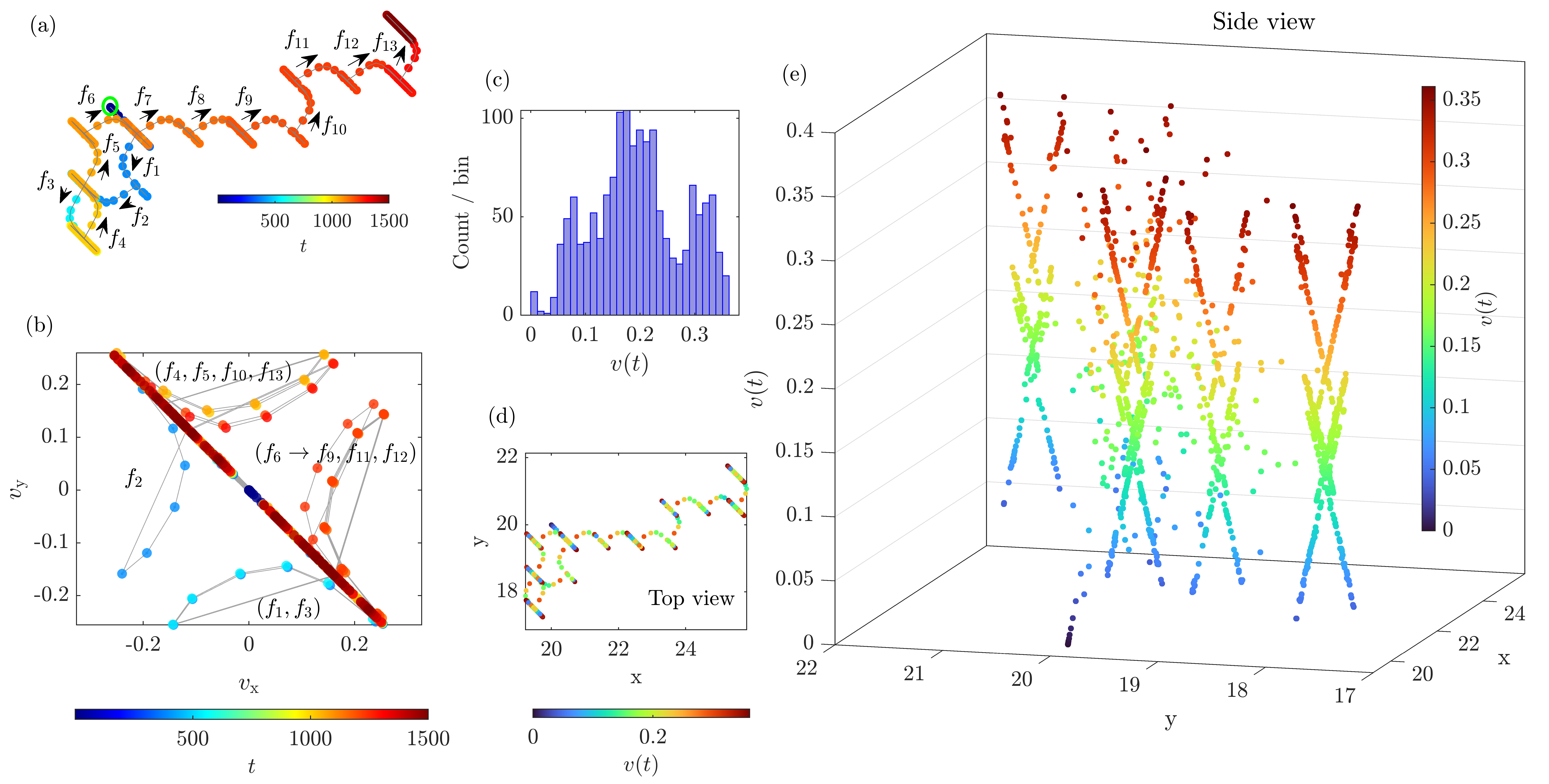} \\
    \vspace{0.45em}
    \includegraphics[width=\textwidth]{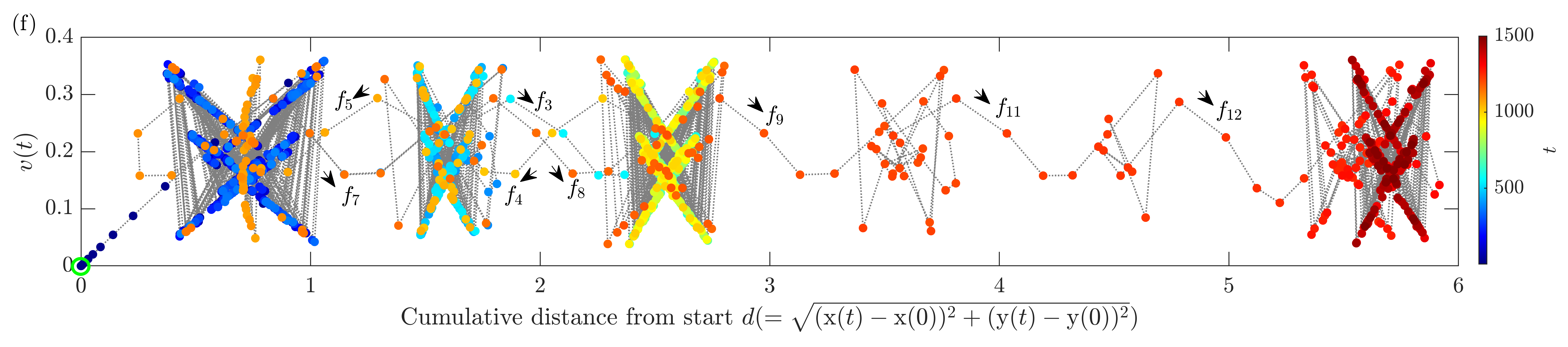}

    \caption{
\textbf{Phase-locked trajectory and burst geometry under memory-mediated feedback.}
(a) Time-colored real-space trajectory at the coherence point \( (E, \eta) = (7.2, 5.1) \), with labeled flights \( f_1 \)–\( f_{13} \) connecting spatially recurring memory towers—zones of speed suppression and directional redirection. 
(b) Velocity-phase portrait \( (v_{\mathrm{x}}, v_{\mathrm{y}}) \) colored by time, showing discrete lobes reflecting reorientation modes constrained by memory gradients.
(c) Speed histogram revealing a tri-modal structure: high-speed bursts and low-speed stalls at tower edges, and intermediate speeds during tower traversal.
(d) Top-view spatial projection colored by speed, highlighting burst–trap alternation: arcs of rapid exit followed by deceleration en route to the next tower.
(e) 3D trajectory \( (x, y, v(t)) \) visualizes vertical towers formed by repeated speed cycling within memory wells—sites of constructive memory overlap and internal feedback locking.
(f) Speed vs cumulative displacement \( d \) from the start reveals recurrent X-shaped motifs at tower locations. Each burst shows acceleration at entry, slowdown near the core, and re-acceleration at exit.
}

    \label{fig:trajectory_structure_combined}
\end{figure*}

\begin{figure*}[!htbp]
    \centering
    \includegraphics[width=0.98\textwidth]{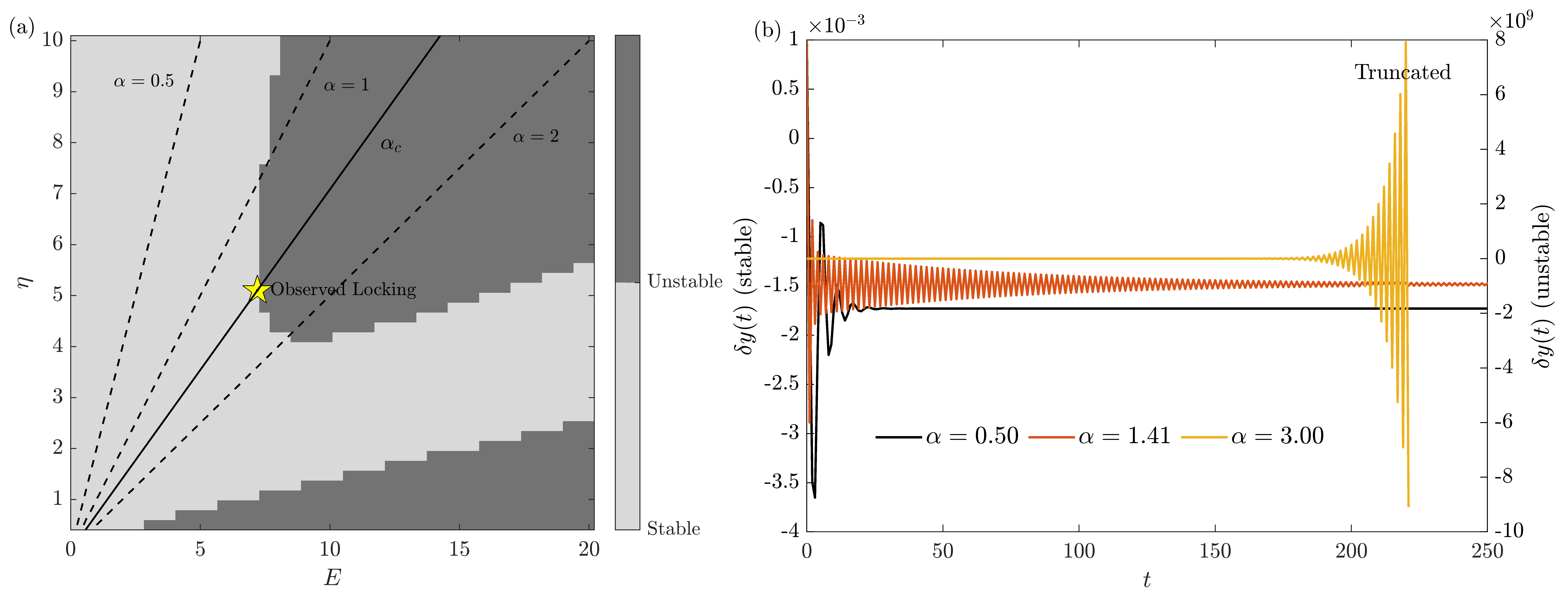}
    \caption{
\textbf{Stability analysis reveals bifurcation underlying phase-locked dynamics.}
(a) Linear stability phase diagram in the \( (E, \eta) \) parameter space. Blue indicates stability of straight-line motion to small transverse perturbations, while red denotes instability leading to locked or curved trajectories. The solid black line represents the critical feedback threshold \( \alpha_c \sim E/\eta \), derived analytically, and closely tracks the numerically observed transition. Dashed lines mark other constant-\( \alpha \) contours. The yellow star indicates the empirically observed coherence point where transfer entropy and feedback alignment peak.
(b) Evolution of transverse deviation \( \delta y(t) \) under three feedback strengths. For \( \alpha = 0.5 \), perturbations decay; at \( \alpha = \alpha_c \approx 1.41 \), damped oscillations mark marginal stability; for \( \alpha = 3.0 \), deviations diverge exponentially, indicating instability. Together, these results show that coherence arises via a symmetry-breaking bifurcation when feedback curvature overwhelms self-memory damping.
}

    \label{fig:stability_analysis}
\end{figure*}

\subsection{Geometry Analogy: Memory Curves and Fracture Paths}

Strikingly, the phase-locked trajectories observed in our simulations resemble the curvilinear fracture paths seen in gypsum under uniaxial compression~\cite{Bobet1998}. In those experiments, coalescing cracks—especially between non-overlapping flaws—trace stabilized arcs that visually echo the curvature-locked motion of our memory-coupled particle. Just as wing cracks initiate at flaw tips and propagate under sustained stress, the particle advances along self-generated memory gradients, producing directionally reinforced, rhythmically curved motion. Although the underlying mechanisms differ—mechanical rupture versus feedback-mediated navigation—the emergent geometries suggest a shared organizing principle: persistent directional coherence shaped by distributed history and local constraints. This analogy hints at a deeper correspondence between fracture coalescence in brittle media and memory-guided diffusion in soft active systems, potentially unifying them within a broader class of structure-forming processes driven by constrained reinforcement and anisotropic feedback.

\subsection{Burst Dynamics and Spectral Resonance}

While previous analyses identified \emph{where} coherence emerges, we now explore \emph{how} it unfolds in time. Within the coherence regime, the particle’s motion is not uniformly smooth, but punctuated by discrete bursts—short-lived accelerations synchronized with energy exchange and surface activity. As shown in Figure~\ref{fig:burst_psd}(a--c), these bursts are highly correlated events: peaks in particle speed align precisely with surges in energy injection \( \mathcal{I}(t) \), dissipation \( \mathcal{D}(t) \), and local surface deformation \( S(x(t), y(t)) \). Rather than continuously drifting through its memory field, the particle enters a rhythmic mode of motion: episodic re-engagements with its own past, triggering rapid feedback loops amplified by constructive gradient overlap.

To quantify this resonance, we examine the spectral structure of the system (Figure~\ref{fig:burst_psd}d). The power spectral densities (PSDs) of both the particle’s speed and the local surface signal show marked enhancement in the low-frequency band \( f \sim 0.1{-}0.3\), diverging significantly from a white noise baseline. This band matches the characteristic recurrence time of burst cycles—slow enough to accumulate imprint memory, yet fast enough to sustain self-feedback.

More telling is the cross-spectral coherence (bottom inset), which exhibits a pronounced dip near \( f \sim 0.2\), indicating a loss of synchronization between surface and speed fluctuations at this characteristic frequency. The phase lag spectrum (top inset) complements this behavior: the deviation from flatness suggests a frequency-dependent delay between surface buildup and velocity response. Together, these features reveal that coherence is not continuous, but organized in bursts, with feedback-mediated rhythms modulating the timing of motion and memory.

\subsection{Phase-Space Geometry of Coherent Motion}
We now trace the internal anatomy of this coherence through a representative trajectory in the locking regime \( (E, \eta) = (7.2, 5.1) \), shown in Figure~\ref{fig:trajectory_structure_combined}. What initially appears as a wandering path resolves into a modular sequence of thirteen flights \( f_1 \)–\( f_{13} \), each linking two localized regions of reduced velocity—\emph{memory towers} where spatial overlap with past trajectories accumulates. These zones trap the particle momentarily and redirect its path via the evolving memory gradient. The top-view trajectory in panel~(a), color-coded by time, reveals how the particle repeatedly accelerates out of one tower, traverses space, and decelerates before entering another—establishing a burst–trap cycle.

Panel~(d), showing the same trajectory colored by speed, captures this dynamic explicitly: rapid acceleration marks tower exit, followed by smooth deceleration as the particle approaches the next site of memory convergence. These arcs form the kinematic scaffolding of coherence: segments of deterministic reorientation shaped by an internal, spatially structured memory field.

The velocity phase-space portrait in panel~(b) reveals further structure. Time-colored loops in the \( (v_{\mathrm{x}}, v_{\mathrm{y}}) \) plane form distinct lobes, each corresponding to a subset of flights. This partitioning indicates that the particle accesses a finite repertoire of directional modes—phase-locked responses constrained by memory geometry rather than random fluctuations.

Panel~(c) confirms this in the statistical domain: the speed histogram exhibits a tri-modal distribution. Outer peaks correspond to fast exits and slow stalls at the edges of memory towers, while the central peak reflects mid-tower and inter-tower traversal. Panel~(f), plotting speed versus cumulative displacement, unveils these cycles as a set of repeated “X”-like motifs: bursts of acceleration, mid-flight slowdown, and renewed speed near tower entry. These patterns are not incidental—they trace the underlying rhythm of feedback-mediated navigation.

Panel~(e) offers a 3D phase-space perspective \( (x, y, v(t)) \), exposing vertically modulated columns—\emph{memory towers}—where the particle returns repeatedly at varying speeds. These columns are not traversals, but local vertical modulations: the particle repeatedly enters, accelerates, and stalls at specific spatial sites. This resonance is governed by the gradient force term in Eq.~\eqref{eq:r_operator},
\[
\nabla S(\mathbf{r}(t), t) = A \int_0^t \Theta_s(t - \tau)\, \nabla G_\sigma(\mathbf{r}(t) - \mathbf{r}(\tau))\, d\tau,
\]
which performs a convolution over past imprints. When \( \mathbf{r}(t) \approx \mathbf{r}(\tau) \) for many \( \tau \), constructive interference intensifies the gradient force, triggering a burst of acceleration. These bursts are thus not stochastic: they are self-induced echoes from the particle’s own past.
\subsection{Numerical Validation of Stability Landscape}

To explain the onset of this behavior, we now turn to the dynamical origin of coherence. Specifically, we analyze whether a particle moving at constant velocity \( \mathbf{v}_0 \) remains stable under a small transverse perturbation \( \delta y(t) \). As derived in Eq.~\eqref{eq:perturbation_equation}, the dynamics of \( \delta y(t) \) involve a competition between damping from the intrinsic kernel \( \Theta_m(t) \) and destabilization from curvature in the memory field, encoded in \( \Lambda(t) \). The system becomes unstable when feedback curvature amplifies transverse deviations faster than self-memory can suppress them.

Figure~\ref{fig:stability_analysis} illustrates this transition. Panel~(a) shows a numerically computed phase diagram in the \( (E, \eta) \) plane, classifying each parameter pair as stable or unstable depending on whether \( \delta y(t) \) grows or decays. The boundary aligns closely with the analytical prediction \( \alpha_c = E/\eta \sim 1/\tau_s \), marked by a solid curve. A yellow star marks the empirically observed coherence point \( (E, \eta) = (7.2, 5.1) \), confirming that phase-locking emerges precisely at the edge of transverse instability.

Panel~(b) illustrates the transition directly by plotting \( \delta y(t) \) for three representative feedback rates. Below threshold (\( \alpha = 0.5 \)), the perturbation decays; at threshold (\( \alpha \approx 1.41 \)), oscillatory relaxation indicates marginal stability; and above threshold (\( \alpha = 3.0 \)), the deviation grows exponentially (truncated at \( 10^{10} \)). This divergence signals a dynamical bifurcation: directional persistence becomes unstable, and the system transitions into a regime of curvature and locking.

Together, these results show that coherence in memory-coupled systems is not a fragile emergent state, but a bifurcation-driven outcome of internal feedback dynamics. When gradient-induced feedback overcomes self-damping, motion reorganizes into structured, resonant trajectories. Coherence, in this framework, is not a product of tuning—but an inevitable phase of self-organized memory dynamics.

\section*{General Discussion and Conclusion}

This minimal model, though simple in construction, reveals a powerful design principle: coherence does not require control—it emerges from coupling. When memory and feedback align, structure self-organizes from within, without the need for external constraint. By embedding history directly into the substrate and allowing it to dynamically influence motion, the system generates order not through external guidance, but through the internal logic of interaction with its own past.

Such mechanisms are not limited to idealized models. In real-world systems—ranging from cytoskeletal rearrangements and morphogenetic flows to active gels, soft robotics, and collective cellular behavior—feedback loops mediated by spatial memory play a central role in organizing complex motion. Our model isolates and abstracts this feedback–memory interplay into its most essential form, allowing us to identify the underlying dynamical signatures and thresholds for coherent behavior.

As systems grow in complexity—featuring multiple interacting agents, adaptive substrates, and distributed memory architectures—these principles offer more than a lens for interpretation. They provide a design rule: coherence can be engineered not by enforcing order, but by tuning the interaction between present dynamics and past imprints. What begins here as a minimal realization of the Coupled Memory Graph Process (CMGP) may ultimately serve as a generative blueprint for constructing self-organizing systems—where motion and memory are not separate entities, but co-orchestrators of emergent structure. Such systems do not require centralized control or fine-tuning; they remember, respond, and reorganize, turning disorder into rhythm through the dynamical logic of their own unfolding history.

\begin{acknowledgments}

The author gratefully acknowledges Prof. Matthias Weiss (University of Bayreuth, Germany) and Prof. Manoj Kumbhakar (BARC, India) for their valuable encouragement, insightful discussions, and continued support. This work did not receive any specific funding but was carried out with institutional support from the Bhabha Atomic Research Centre, Department of Atomic Energy, India and the University of Bayreuth, Germany.

\end{acknowledgments}

\bibliographystyle{unsrt}      

\bibliography{refs}  

\end{document}